\documentclass[aps,prb,fleqn,12pt,showpacs]{revtex4}
\usepackage{epsfig,color}
\usepackage{graphicx}
\usepackage{pdfpages}
\usepackage{braket}
\usepackage{amssymb,amsmath}
\usepackage{amsfonts} %txfonts}
\usepackage{hyperref} 
\begin{document}
\title{Pseudo-spin rotation symmetry breaking by \\ Coulomb interaction terms in spin-orbit coupled systems}
\author{Shubhajyoti Mohapatra}
%\affiliation{Department of Physics, Indian Institute of Technology, Kanpur - 208016, India}
\author{Avinash Singh}
\email{avinas@iitk.ac.in}
\affiliation{Department of Physics, Indian Institute of Technology, Kanpur - 208016, India}
\date{\today} 
\begin{abstract} 
By transforming from the pure-spin-orbital ($t_{\rm 2g}$) basis to the spin-orbital entangled pseudo-spin-orbital basis, the pseudo-spin rotation symmetry of the different Coulomb interaction terms is investigated under SU(2) transformation in pseudo-spin space. While the Hubbard and density interaction terms are invariant, the Hund's coupling and pair-hopping interaction terms explicitly break pseudo-spin rotation symmetry systematically. The form of the symmetry-breaking terms obtained from the transformation of the Coulomb interaction terms accounts for the easy $x$-$y$ plane anisotropy and magnon gap for the out-of-plane mode, highlighting the importance of mixing with the nominally non-magnetic $J$=3/2 sector, and providing a physically transparent approach for investigating magnetic ordering and anisotropy effects in perovskite ($\rm Sr_2 Ir O_4$) and other $d^5$ pseudo-spin compounds. 

\end{abstract}
% Key words: Coulomb Interaction Terms, Pseudo-Spin-Orbital Basis, Spin-Orbit coupling, Spin Wave Gap
\maketitle
\newpage
\section{Introduction}
Arising from a novel interplay between crystal field, spin-orbit coupling (SOC) and intermediate-strength Coulomb interactions, the emergent quantum states which essentially determine the electronic and magnetic properties of the iridium based transition-metal oxides involve correlated motion of electrons in spin-orbital entangled states.\cite{krempa_AR_2014,rau_AR_2016,bertinshaw_AR_2018} In the spin-orbit Mott insulator $\rm Sr_2IrO_4$ with $d^5$ configuration, electronic states near the Fermi energy have dominantly $J$=1/2 character, and important magnetic properties such as in-plane canted antiferromagnetic (AFM) order and magnon excitations have been extensively discussed in terms of the effectively single pseudo orbital ($J$=1/2) picture.\cite{kim_PRL_2008,kim_Science_2009,watanabe_PRL_2010,kim1_PRL_2012} Finite-interaction and finite-SOC effects are responsible for the strong zone-boundary magnon dispersion measured in resonant inelastic X-ray scattering (RIXS) studies, highlighting the observable effect of mixing between $J$=1/2 and 3/2 sectors.\cite{iridate_one} 

The Dzyaloshinskii-Moriya (DM) and pseudo-dipolar (PD) anisotropic interactions in the $J=1/2$ sector, although weakly affected by the tetragonal splitting,\cite{kim_PRL_2012} are not the source of true anisotropy in $\rm Sr_2IrO_4$, as they yield pseudo-spin canting with no magnon gap due to compensation. True anisotropy has been ascribed to the Hund's coupling term ($J_{\rm H}$) using strong-coupling expansion (including virtual excitations to $J$=3/2 states) and numerical self-consistent calculation.\cite{jackeli_PRL_2009,igarashi_PRB_2013,perkins_PRB_2014,vale_PRB_2015,igarashi_JPSJ_2014} While Coulomb interactions were considered within the pure-spin-orbital basis ($t_{\rm 2g}$ orbitals, pure spins) in above approaches, their treatment within a pseudo-spin-orbital basis, and the role of weak magnetism in the other two pseudo orbitals ($J$=3/2 sector) on the $J_{\rm H}$-induced easy-plane magnetic anisotropy and magnon gap ($\sim 40$ meV), as measured in recent resonant inelastic X-ray scattering (RIXS) studies,\cite{kim_NATCOMM_2014,pincini_PRB_2017,porras_PRB_2019} have not been elucidated. Furthermore, the pseudo-spin-orbital based approach can allow for a unified study of both intra-orbital (magnon) and inter-orbital (spin-orbit exciton) excitations within a single formalism.

%has been studied recently including Coulomb interactions within a pseudo-spin-orbital based approach, which allows for a unified calculation of both intra-orbital (magnon) and inter-orbital (spin-orbit exciton) excitations within a single formalism.\cite{iridate_two}

Magnetic anisotropy is generally associated with spin rotation symmetry breaking. Therefore, a general pseudo-spin rotation symmetry analysis of the different Coulomb interaction terms, treating all three pseudo orbitals on the same footing, can provide additional insight into the origin of true magnetic anisotropy in $\rm Sr_2IrO_4$ arising from the interplay of spin-orbital entanglement and Coulomb interaction. Due to the spin-orbital entanglement, the same pseudo-spin rotation for all three pseudo orbitals ($l=1,2,3$) corresponds to different pure-spin rotations for the three pure ($t_{\rm 2g}$) orbitals. This follows directly from the relation $\psi_\mu = \sigma_\mu \sum_l c_{\mu l} \psi_l$ between the fermionic field operators in the pure-spin-orbital basis ($\mu=yz,xz,xy$) and pseudo-spin-orbital basis ($l=1,2,3$), where the Pauli matrices $\sigma_\mu = \sigma_x,\sigma_y,\sigma_z$ corresponding to the three pure orbitals $\mu=yz,xz,xy$. The same SU(2) transformation $\psi_l \rightarrow \psi_l ' = [U]\psi_l$ for all three pseudo orbitals corresponds to different SU(2) transformations $\psi_\mu \rightarrow \psi_\mu '=[U_\mu]\psi_\mu$ for the three pure orbitals due to the orbital dependence of $[U_\mu]=\sigma_\mu [U] \sigma_\mu$.

Therefore, the question of how the different Coulomb interaction terms transform under same pseudo-spin rotation for all three pseudo orbitals assumes importance. In other words, while all Coulomb interaction terms are invariant under same pure-spin rotation for all three pure orbitals, does this invariance hold under same pseudo-spin rotation in the pseudo-spin-orbital basis? Pseudo-spin rotation symmetry breaking by any Coulomb interaction term would imply true magnetic anisotropy and gapped magnon spectrum. 

In this paper, we will show that while the Hubbard $(U)$ and density $(U')$ interaction terms do preserve pseudo-spin rotation symmetry, the Hund's coupling ($J_{\rm H}$) and pair-hopping ($J_{\rm P}$) interaction terms explicitly break this symmetry systematically. This symmetry breaking results in (on-site) anisotropic interactions between moments in the $J$=1/2 and $J$=3/2 sectors, highlighting the importance of the weak magnetism in the nominally filled $J$=3/2 sector as well as the mixing between the two sectors. Magnetic anisotropy will not survive in the large SOC limit when the two sectors become effectively decoupled. The physically transparent approach for investigating magnetic ordering and anisotropy effects will be illustrated for the perovskite compound $\rm Sr_2 Ir O_4$, where the sign of $J$=3/2 sector moments directly yields easy $x$-$y$ plane anisotropy. 

%As pseudo-spin moments in the $J$=3/2 sector are order of magnitude smaller compared to $J$=1/2 moment, a pseudo-spin-orbital based approach offers significant practical advantage by allowing even smaller anomalous condensates to be dropped in a numerical self-consistent calculation. This allows deeper insight into the subtle effects on magnetic anisotropy as mentioned above. On the other hand, with comparable spin moments for all three real orbitals, a large basis set may need to be employed to include the various condensates in the real-spin-orbital approach.

The structure of this paper is as below. After introducing the transformation between the pure-spin-orbital and pseudo-spin-orbital bases in Sec. II, a symmetry analysis of the individual Coulomb interaction terms is carried out in Sec. III, where the Hund’s coupling and pair hopping interaction terms are shown to explicitly break pseudo-spin rotation symmetry. The transformed Coulomb interaction terms are obtained in Sec. IV, explicitly showing that the symmetry breaking terms account for the easy $x$-$y$ plane anisotropy in $\rm Sr_2 Ir O_4$, as illustrated for the ground state energy in Sec. V, and for magnon excitations and anisotropy gap in Sec. VI. Comparison with an independent $t_{\rm 2g}$ orbital based approach is discussed in Sec. VII, which provides confirmation of the above symmetry analysis and also illustrates the significant simplification achieved in the pseudo-spin-orbital basis. After a critical comparison (Sec. VIII) of different but equivalent approaches for studying magnetic anisotropy in $\rm Sr_2IrO_4$, some conclusions are  finally presented in Sec. IX. 

%recently developed for the $\rm Ca_2RuO_4$ compound based on the real ($t_{2g}$ sector) orbital basis (ref. 23), and extending to the $\rm Sr_2IrO_4$ compound with electron filling $n=5$, we have explicitly confirmed that the easy-plane anisotropy is (i) due to $J_{\rm H}$ and $J_{\rm P}$, and (ii) not due to the tetragonal distortion term $\epsilon_{xy}$. Furthermore, comparison of magnetization and density values in the real and pseudo orbital bases illustrates the significant simplification achieved in the pseudo orbital basis. (Sec. VII)

%and their behavior under SU(2) transformation in pseudo-spin space are studied
%(keeping the Hubbard, Hund's coupling and density interaction terms), 

\section{Pseudo-spin-orbital basis}

\begin{figure}
\vspace*{-10mm}
\hspace*{0mm}
\psfig{figure=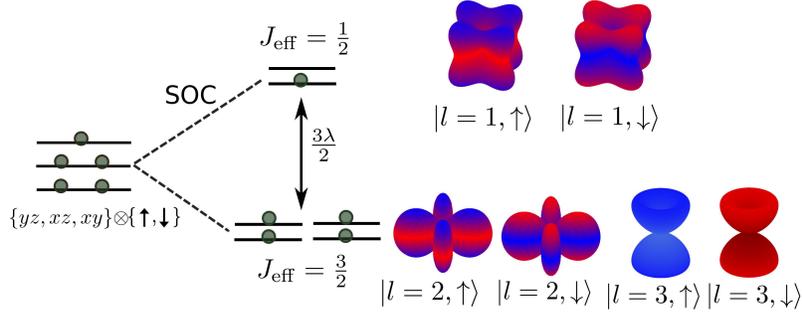,angle=0,width=105mm}
\vspace{-0mm}
\caption{The pseudo-spin-orbital energy level scheme for the three Kramers pairs along with their orbital shapes. The colors represent the weights of pure spins ($\uparrow$-red, $\downarrow$-blue) in each pair.} 
\label{schematic}
\end{figure}

Due to large crystal-field splitting ($\sim$3 eV) in the $\rm Ir O_6$ octahedra, low-energy physics in $d^5$ iridates is effectively described by projecting out the empty ${\rm e_g}$ levels which are well above the $t_{\rm 2g}$ levels. Spin-orbit coupling (SOC) further splits the $t_{\rm 2g}$ states into $J$=1/2 doublet ($m_J=\pm 1/2$) and $J$=3/2 quartet ($m_J=\pm 1/2,\pm 3/2$), with an energy gap of $3\lambda/2$ (Fig. \ref{schematic}). Nominally, four of the five electrons fill the $J$=3/2 states, leaving one electron for the $J$=1/2 sector, rendering it magnetically active in the ground state. 

In the ($yz$$\downarrow$,$xz$$\downarrow$,$xy$$\uparrow||\; yz$$\uparrow$,$xz$$\uparrow$,$xy$$\downarrow$) basis, the SOC Hamiltonian becomes block diagonal with two [$3\times 3$] blocks. There is no mixing between the ($\downarrow,\downarrow,\uparrow$) and ($\uparrow,\uparrow,\downarrow$) sectors, which can therefore be treated as pseudo spins $\uparrow$ and $\downarrow$ for the three doubly degenerate eigenstates (pseudo orbitals).\cite{watanabe_PRL_2010} Corresponding to the three Kramers pairs $|J,m_j \rangle$ above, the pseudo-spin-orbital basis states $|l,\tau \rangle$ for the three {\em pseudo orbitals} ($l=1,2,3$), with {\em pseudo spins} ($\tau= \uparrow,\downarrow$) each, have the form:
\begin{eqnarray}
\ket{l=1, \tau= \sigma} &=& \Ket{\frac{1}{2},\pm\frac{1}{2}} = \left [\Ket{yz,\bar{\sigma}} \pm i \Ket{xz,\bar{\sigma}} \pm \Ket{xy,\sigma}\right ] / \sqrt{3} \nonumber \\
\ket{l=2, \tau= \sigma} &=& \Ket{\frac{3}{2},\pm\frac{1}{2}} = \left [\Ket{yz,\bar{\sigma}} \pm i \Ket{xz,\bar{\sigma}} \mp 2 \Ket{xy,\sigma} \right ] / \sqrt{6} \nonumber \\
\ket{l=3, \tau= \bar{\sigma}} &=& \Ket{\frac{3}{2},\pm\frac{3}{2}} = \left [\Ket{yz,\sigma} \pm i \Ket{xz,\sigma}\right ] / \sqrt{2} 
\label{jmbasis}
\end{eqnarray}
where $\Ket{yz,\sigma}$, $\Ket{xz,\sigma}$, $\Ket{xy,\sigma}$ are the t$_{2g}$ basis states and the signs $\pm$ correspond to spins $\sigma = \uparrow/\downarrow$. The coherent superposition of different-symmetry $t_{\rm 2g}$ orbitals, with opposite spin polarization between $xz$/$yz$ and $xy$ levels implies spin-orbital entanglement, and also imparts unique extended 3D shape to the pseudo-orbitals $l=1, 2, 3$, as shown in Fig \ref{schematic}. The pseudo-spin dynamics in iridate heterostructures are gaining interest as their magnetic properties are much more sensitive to structural distortion compared to pure spin systems due to spin-orbital entanglement.\cite{meyers_SREP_2019,mohapatra_PRB_2019}

Taking the conjugate to express the above basis transformation in terms of the $\langle l,\tau|$ and $\langle \mu,\sigma|$ states, and rewriting in terms of the corresponding fermionic field operators: 
\begin{equation}
\psi_l = \left ( \begin{array}{c} a_{l\uparrow} \\ a_{l\downarrow} \end{array} \right ) 
\;\;\; {\rm and} \;\;\; 
\psi_\mu = \left ( \begin{array}{c} a_{\mu\uparrow} \\ a_{\mu\downarrow} \end{array} \right ) 
\end{equation}
involving the annihilation operators for the pseudo orbitals ($l=1,2,3$, $\tau = \uparrow,\downarrow$) and the $t_{2g}$ orbitals ($\mu=yz,xz,xy$, $\sigma = \uparrow,\downarrow$), we obtain (using Pauli matrices):
\begin{eqnarray}
\psi_{1} & = & \frac{1}{\sqrt{3}} \left [\sigma_x \psi_{yz} + \sigma_y \psi_{xz} + \sigma_z \psi_{xy} \right ] \nonumber \\
\psi_{2} & = & \frac{1}{\sqrt{6}} \left [\sigma_x \psi_{yz} + \sigma_y \psi_{xz} - 2 \sigma_z \psi_{xy} \right ] \nonumber \\
\psi_{3} & = & \frac{1}{\sqrt{2}} \left [\sigma_x \psi_{yz} - \sigma_y \psi_{xz} \right ] .
\end{eqnarray}
Inverting the above transformation yields the $t_{2g}$ basis states represented in terms of the pseudo-spin-orbital basis states:
\begin{eqnarray}
\psi_{yz} & = & \sigma_x \left [ \frac{1}{\sqrt{3}} \psi_1 + \frac{1}{\sqrt{6}} \psi_2 + \frac{1}{\sqrt{2}} \psi_3 \right ] \nonumber \\
\psi_{xz} & = & \sigma_y \left [ \frac{1}{\sqrt{3}} \psi_1 + \frac{1}{\sqrt{6}} \psi_2 - \frac{1}{\sqrt{2}} \psi_3 \right ] \nonumber \\
\psi_{xy} & = & \sigma_z \left [ \frac{1}{\sqrt{3}} \psi_1 - \sqrt{\frac{2}{3}} \psi_2 \right ] .
\label{trans}
\end{eqnarray}
The above equations are convenient for transforming the hopping and Coulomb interaction terms to the pseudo-spin-orbital basis, and can be expressed in the compact form:
\begin{equation} 
\psi_{\mu} = \sigma_\mu \sum_{l=1,2,3} c_{\mu l} \, \psi_l
\label{trans2}
\end{equation}
where $\sigma_\mu = \sigma_x,\sigma_y,\sigma_z$ for the three orbitals $\mu=yz,xz,xy$, respectively, and the (real) transformation coefficients $c_{\mu l}$ are explicitly shown in Eq. (\ref{trans}).

%\subsection*{Effect of tetragonal distortion}
%The transformation coefficients $c_{\mu l}$ in the above equation are simply modified when the tetragonal distortion effect is included by the term $\epsilon_{xy} \psi_{xy}^\dagger \psi_{xy}$ in the $t_{2g}$ basis SOC Hamiltonian, where the tetragonal splitting $\epsilon_{xy}$ is the $xy$ orbital energy offset relative to the degenerate $yz,xz$ orbitals.\cite{boseggia_2014} While energy of the $(J,m_J)=(3/2,3/2)$ pair remains unchanged as this state $(l=3)$ has no $xy$ orbital character (Eq. \ref{jmbasis}), the $(1/2,1/2)$ and $(3/2,1/2)$ pairs are shifted. However, as the spin-orbital entanglement remains unaffected, the tetragonal distortion has no effect on the SU(2) symmetry analysis discussed below. 

%Although the tetragonal splitting $\epsilon_{xy}$ does weakly affect the DM and PD anisotropic interactions in the $J=1/2$ sector,\cite{kim_PRL_2012} it is not the source of true magnetic anisotropy. As discussed in section VII, the $J=1/2$ sector interaction terms extracted from the Hund's coupling and pair-hopping terms individually acquire classical anisotropy when the octahedral cubic symmetry is lifted by the tetragonal distortion, as present in $\rm Sr_2IrO_4$. However, classically anisotropic interaction terms such as $S_{1z} S_{1z}$ thus generated do not yield true magnetic anisotropy.

% Transformation of Coulomb interaction terms and 
\section{Pseudo-spin rotation symmetry breaking}
We consider the on-site Coulomb interaction terms in the $t_{2g}$ basis ($\mu,\nu=yz,xz,xy$):
\begin{eqnarray}
\mathcal{H}_{\rm int} &=& U\sum_{i,\mu}{n_{i\mu\uparrow}n_{i\mu\downarrow}} + U^\prime \sum_{i,\mu < \nu,\sigma} {n_{i\mu\sigma} n_{i\nu\overline{\sigma}}} + (U^\prime - J_{\mathrm H}) \sum_{i,\mu < \nu,\sigma}{n_{i\mu\sigma}n_{i\nu\sigma}} \nonumber\\ 
&+& J_{\mathrm H} \sum_{i,\mu \ne \nu} {a_{i \mu \uparrow}^{\dagger}a_{i \nu\downarrow}^{\dagger}a_{i \mu \downarrow} a_{i \nu \uparrow}} + J_{\mathrm P} \sum_{i,\mu \ne \nu} {a_{i \mu \uparrow}^{\dagger} a_{i \mu\downarrow}^{\dagger}a_{i \nu \downarrow} a_{i \nu \uparrow}} \nonumber\\ 
&=& U\sum_{i,\mu}{n_{i\mu\uparrow}n_{i\mu\downarrow}} + U^{\prime \prime}\sum_{i,\mu<\nu} n_{i\mu} n_{i\nu} - J_{\mathrm H} \sum_{i,\mu \ne \nu} {\bf S}_{i\mu} \cdot {\bf S}_{i\nu} 
+J_{\mathrm P} \sum_{i,\mu \ne \nu} a_{i \mu \uparrow}^{\dagger} a_{i \mu\downarrow}^{\dagger}a_{i \nu \downarrow} a_{i \nu \uparrow} 
\label{inter}
\end{eqnarray} 
including the intra-orbital $(U)$ and inter-orbital $(U')$ density interaction terms, the Hund's coupling term $(J_{\rm H})$, and the pair hopping interaction term $(J_{\rm P}$=$J_{\rm H})$. Here $a_{i\mu\sigma}^{\dagger}$ and $a_{i\mu \sigma}$ are the creation and annihilation operators for site $i$, orbital $\mu$, spin $\sigma=\uparrow ,\downarrow$, the density operator $n_{i\mu\sigma}=a_{i\mu\sigma}^\dagger a_{i\mu\sigma}$, the total density operator $n_{i\mu}=n_{i\mu\uparrow}+n_{i\mu\downarrow}=\psi_{i\mu}^\dagger \psi_{i\mu}$, and $U^{\prime\prime}=U^\prime-J_{\rm H}/2$. All interaction terms above are SU(2) invariant and thus possess pure-spin rotation symmetry. In the following, we consider the transformation of individual Coulomb interaction terms to the pseudo-spin-orbital basis using Eq. (\ref{trans2}), and examine their SU(2) transformation behavior in pseudo-spin space. 

\subsection{Total density operator}
For the total density operator (for site $i$), we obtain using Eq. (\ref{trans2}):
\begin{equation} 
n_{\mu} = \psi_{\mu}^\dagger \psi_{\mu} = \sum_{l,m} c_{\mu l} c_{\mu m} \psi_l^\dagger \psi_m 
\label{tot_dens}
\end{equation}
where we have used $\sigma_\mu^\dagger = \sigma_\mu$ and $\sigma_\mu ^2 = {\bf 1}$. Now, under the SU(2) transformation in pseudo-spin space (same for all three pseudo orbitals $l$):
\begin{equation}
\psi_l \rightarrow \psi_l ' = [U] \psi_l
\label{su2}
\end{equation}
the $\psi_l^\dagger \psi_m$ terms are invariant, and the total density operator is therefore SU(2) invariant. Therefore, the density interaction terms ($U'$) in Eq. (\ref{inter}) and the Hubbard interaction terms ($U$) (using $(n_\uparrow + n_\downarrow)^2 = n_\uparrow + n_\downarrow + 2n_\uparrow n_\downarrow$) are SU(2) invariant and possess spin rotation symmetry in pseudo-spin space. 

\subsection{Pair hopping interaction term}
For the pair hopping interaction term (for site $i$), we obtain:
\begin{eqnarray}
J_{\rm H} \; a_{\mu \uparrow}^\dagger a_{\mu \downarrow}^\dagger a_{\nu \downarrow} a_{\nu \uparrow} &=& 
J_{\rm H} \; a_{\mu \uparrow}^\dagger a_{\nu \uparrow} a_{\mu \downarrow}^\dagger a_{\nu \downarrow} = \frac{J_{\rm H}}{2} \Big(a_{\mu \uparrow}^\dagger a_{\nu \uparrow} + a_{\mu \downarrow}^\dagger a_{\nu \downarrow} \Big) \Big(a_{\mu \uparrow}^\dagger a_{\nu \uparrow} +  a_{\mu \downarrow}^\dagger a_{\nu \downarrow} \Big) \nonumber \\
&=& \frac{J_{\rm H}}{2} \left ( \psi_\mu^\dagger \psi_\nu \right )^2,
\label{phterm}
\end{eqnarray}
which is SU(2) invariant and possesses spin-rotation symmetry in pure-spin space. However, SU(2) invariance is lost in pseudo-spin space, as shown below. Again, using Eq. (\ref{trans2}) to transform to the pseudo-spin-orbital basis, we obtain: 
\begin{equation}
\psi_\mu^\dagger \psi_\nu = \sum_{l,m} c_{\mu l} c_{\nu m} \; \psi_l^\dagger (\sigma_x \sigma_y) \psi_m 
= \sum_{l,m} c_{\mu l} c_{\nu m} \; \psi_l^\dagger (i\sigma_z) \psi_m ,
\label{ph_dens}
\end{equation}
where we have taken $\mu=yz$, $\nu=xz$ to illustrate the operations with Pauli matrices. Now, under the SU(2) transformation in pseudo-spin space (Eq. \ref{su2}), the last term in Eq. (\ref{ph_dens}): 
\begin{equation}
\psi_l^\dagger \left(i\sigma_z \right) \psi_m \rightarrow \psi_l^\dagger \left[ U \right]^\dagger \left(i\sigma_z \right) \left[U\right] \psi_m \ne \psi_l^\dagger \left( i\sigma_z \right) \psi_m ,
\end{equation}
showing that $\psi_\mu^\dagger \psi_\nu$ is not SU(2) invariant. The pair hopping interaction term therefore explicitly breaks pseudo-spin rotation symmetry.

\subsection{Hund's coupling term}
For this term involving the pure-spin rotationally symmetric interaction ${\bf S}_{i\mu}\cdot{\bf S}_{i\nu}$, we consider the spin density operator (for site $i$), and obtain using Eq. (\ref{trans2}):
\begin{equation}
2{\bf S}_\mu = \psi_\mu ^\dagger \makebox{\boldmath $\sigma$} \psi_\mu = \sum_{lm} c_{\mu l} c_{\mu m} \psi_l ^\dagger (\sigma_\mu \makebox{\boldmath $\sigma$} \sigma_\mu) \psi_m
\label{hc_one}
\end{equation}
which transforms under the SU(2) transformation (Eq. \ref{su2}) to:
\begin{equation}
2{\bf S}_\mu \rightarrow 2{\bf S}_\mu ' = \psi_\mu '^{\dagger} \makebox{\boldmath $\sigma$} \psi_\mu ' = \sum_{lm} c_{\mu l} c_{\mu m} \psi_l ^\dagger \left ( [U]^\dagger \sigma_\mu \makebox{\boldmath $\sigma$} \sigma_\mu [U] \right ) \psi_m
\end{equation}

We now consider the term in brackets above for the case $\sigma_\mu = \sigma_x$ ($yz$ orbital) and represent it in terms of a rotation operation in spin space:
\begin{equation}
[U]^\dagger \sigma_x \left ( \begin{array}{c} \sigma_x \\ \sigma_y \\ \sigma_z \end{array} \right ) \sigma_x [U] = [U]^\dagger \left ( \begin{array}{r} \sigma_x \\ -\sigma_y \\ -\sigma_z \end{array} \right ) [U] = \left ( \begin{array}{r} \sigma_x ' \\ -\sigma_y ' \\ -\sigma_z '\end{array} \right ) = R_x(\pi) R(U) \left ( \begin{array}{c} \sigma_x \\ \sigma_y \\ \sigma_z \end{array} \right )
\end{equation}
where 
\begin{equation}
R(U) \left ( \begin{array}{c} \sigma_x \\ \sigma_y \\ \sigma_z \end{array} \right ) = 
\left ( \begin{array}{r} \sigma_x ' \\ \sigma_y ' \\ \sigma_z '\end{array} \right ) = 
[U]^\dagger \left ( \begin{array}{c} \sigma_x \\ \sigma_y \\ \sigma_z \end{array} \right ) [U]
\end{equation}
shows the spin rotation by the rotation matrix $R(U)$ corresponding to the SU(2) transformation $[U]$, and 
\begin{equation}
R_x(\pi) = \left ( \begin{array}{ccc} 1 & 0 & 0 \\ 0 & -1 & 0 \\ 0 & 0 & -1 \end{array} \right )
\label{r_matrix}
\end{equation}
is the rotation matrix corresponding to $\pi$ rotation about the $x$ axis.

Similarly, for the ${\bf S}_\nu$ operator with $\nu=y$ ($xz$ orbital), we will obtain the product $R_y(\pi) R(U)$. Therefore, the ${\bf S}_\mu\cdot{\bf S}_\nu$ interaction term will yield the matrix product:
\begin{equation}
[\widetilde{R_x(\pi) R(U)}] R_y(\pi) R(U) = \widetilde{R}(U) R_x(\pi) R_y(\pi) R(U) 
\end{equation}
where $\widetilde{R}(U)$ is the transpose of $R(U)$ and we have used $\widetilde{R}_x(\pi) = R_x(\pi)$ for the diagonal matrix. Finally, since
\begin{equation}
\widetilde{R}(U) R_x(\pi) R_y(\pi) R(U) \ne R_x(\pi) R_y(\pi) 
\end{equation}
as $R_x(\pi) R_y(\pi) \ne {\bf 1}$, the Hund's coupling term ${\bf S}_\mu\cdot{\bf S}_\nu$ is not pseudo-spin SU(2) invariant and therefore does not possess pseudo-spin rotation symmetry.

The above symmetry analysis shows that the Hund's coupling and pair hopping interaction terms explicitly break pseudo-spin rotation symmetry. It is important to note here that the spin rotation symmetry is broken systematically. In other words, it is broken for each term in ${\bf S}_\mu\cdot {\bf S}_\nu$ involving the summations over $(l,m)$ and $(l',m')$ for ${\bf S}_\mu$ and ${\bf S}_\nu$ in Eq. (\ref{hc_one}), and similarly for the pair hopping interaction term in Eq. (\ref{ph_dens}).  

\section{Transformed Coulomb interaction terms}

In the following, we illustrate the transformation of the different Coulomb interaction terms to the pseudo-spin-orbital basis, starting with $l$=1 ($J$=1/2) sector of the intra-pseudo-orbital interaction terms. Similar transformation to the $J$ basis has been discussed recently, focussing only on the density interaction terms.\cite{martins_JPCM_2017} Considering first the pair-hopping interaction term (Eq. \ref{phterm}), and retaining only the $l$=$m$=$l'$=$m'$=1 terms (indicated by $\leadsto$ below), we obtain:
\begin{eqnarray}
& &\frac{J_{\rm H}}{2}\sum_{\mu\ne\nu} (\psi_\mu^\dagger \psi_\nu)^2 = \frac{J_{\rm H}}{2}\sum_{\mu\ne\nu} \left [\sum_{l,m} c_{\mu l} c_{\nu m} \psi_l^\dagger (\sigma_\mu \sigma_\nu) \psi_m \right ] \left [\sum_{l',m'} c_{\mu l'} c_{\nu m'} \psi_{l'}^\dagger (\sigma_\mu \sigma_\nu) \psi_{m'} \right ] \nonumber \\
& \leadsto & J_{\rm H} \left [ c_{yz,1}^2 c_{xz,1}^2 \left \{\psi_1^\dagger (i\sigma_z)\psi_1 \right \}^2 + c_{xz,1}^2 c_{xy,1}^2 \left \{ \psi_1^\dagger (i\sigma_x)\psi_1 \right \}^2 + c_{xy,1}^2 c_{yz,1}^2 \left \{ \psi_1^\dagger (i\sigma_y)\psi_1 \right \}^2 \right ] \nonumber \\
&=&-4J_{\rm H} \left [ c_{yz,1}^2 c_{xz,1}^2 \left \{S_{1z}^2\right \} + c_{xz,1}^2 c_{xy,1}^2 \left \{S_{1x}^2\right \} + c_{xy,1}^2 c_{yz,1}^2 \left \{S_{1y}^2\right \}\right ] = -\frac{4J_{\rm H}}{9} {\bf S}_1 \cdot {\bf S}_1  
\label{phterm_2}
\end{eqnarray}

%It should be noted that true magnetic anisotropy does not arise even if classically anisotropic terms such as $S_{1z}S_{1z}$ are present, as when the cubic symmetry is lifted by the tetragonal distortion. This well known property of $S=1/2$ quantum spin operators $S_\mu=(1/2)\psi^\dagger \sigma_\mu \psi$ is discussed in the next section focussing on tetragonal distortion effects.

Similarly, for the Hund's coupling term in Eq. (\ref{inter}), we obtain:
\begin{eqnarray}
-2J_{\rm H} \sum_{\mu<\nu} {\bf S}_\mu . {\bf S}_\nu 
&=& -\frac{J_{\rm H}}{2} \sum_{\mu<\nu} \left [\sum_{l,m} c_{\mu l} c_{\mu m} \psi_l^\dagger (\sigma_\mu \makebox{\boldmath $\sigma$} \sigma_\mu) \psi_m \right ] . \left [\sum_{l',m'} c_{\nu l'} c_{\nu m'} \psi_{l'}^\dagger (\sigma_\nu \makebox{\boldmath $\sigma$} \sigma_\nu) \psi_{m'} \right ] \nonumber \\
&\leadsto & -\frac{J_{\rm H}}{2} \sum_{\mu<\nu} \left [c_{\mu 1}^2 c_{\nu 1}^2 \left \{\psi_1^\dagger (R_\mu \makebox{\boldmath $\sigma$}) \psi_1 \right \}. \left \{\psi_1^\dagger (R_\nu \makebox{\boldmath $\sigma$}) \psi_1 \right \} \right ] \nonumber \\
&=&-2J_{\rm H} \sum_{\mu<\nu} \left [c_{\mu 1}^2 c_{\nu 1}^2 (S_{1x} \;\; S_{1y} \;\; S_{1z}) \widetilde{R}_\mu R_\nu \left (\begin{array}{c} S_{1x} \\ S_{1y} \\ S_{1z} \end{array} \right ) \right ] = \frac{2J_{\rm H}}{9} {\bf S}_1 \cdot {\bf S}_1
%&=& -\frac{2J_{\rm H}}{9} \left [(S_{1x} \; S_{1y} \; S_{1z}) (R_x R_y + R_y R_z + R_z R_x) \left ( \begin{array}{c} S_{1x} \\ S_{1y} \\ S_{1z} \end{array} \right ) \right ] \nonumber \\
%&=& \frac{2J_{\rm H}}{9} {\bf S}_1 . {\bf S}_1  
\label{jhterm_2}
\end{eqnarray}
as the product $c_{\mu 1}^2 c_{\nu 1}^2$ is identical for all three orbital pairs, and for the rotation matrices (Eq. \ref{r_matrix}), we have $R_x R_y + R_y R_z + R_z R_x=-{\bf 1}$. Although the pair-hopping and Hund's coupling interaction terms generally break pseudo-spin rotation symmetry, within the magnetically active $l$=1 ($J$=1/2) sector, they individually yield classically isotropic terms of the form ${\bf S}_1\cdot{\bf S}_1$. This follows from the special symmetry within this sector, as reflected by the identical coefficients $c_{\mu l}$ for all three orbitals $\mu=yz,xz,xy$ for $l=1$ (Eq. \ref{trans}).

Finally, from the remaining $J_{\rm H}$ term (in the $U^{\prime\prime}$ term of Eq. \ref{inter}), again retaining only the $l=m=l'=m'=1$ term, we obtain (using Eq. \ref{tot_dens} for site $i$):
\begin{eqnarray}
-\frac{J_{\rm H}}{2} \sum_{\mu<\nu} n_{\mu} n_{\nu}  
&=& -\frac{J_{\rm H}}{2} \sum_{\mu<\nu} \left ( \sum_{l,m} c_{\mu l} c_{\mu m} \psi_l^\dagger  \psi_m \right ) \left ( \sum_{l',m'} c_{\nu l'} c_{\nu m'} \psi_{l'}^\dagger \psi_{m'} \right ) \nonumber \\
&\leadsto & -\frac{J_{\rm H}}{2} \sum_{\mu<\nu} c_{\mu 1}^2 c_{\nu 1}^2 (\psi_1^\dagger \psi_1)^2
= -\frac{J_{\rm H}}{2} \sum_{\mu<\nu} c_{\mu 1}^2 c_{\nu 1}^2 (n_{1\uparrow} + n_{1\downarrow})
(n_{1\uparrow} + n_{1\downarrow}) \nonumber \\ 
& = & \frac{2J_{\rm H}}{9} {\bf S}_1\cdot{\bf S}_1 - \frac{J_{\rm H}}{3} n_1
\label{jhterm_3}
\end{eqnarray}
where we have used $n_{1\sigma}^2=n_{1\sigma}$ and $2n_{1\uparrow}n_{1\downarrow}=n_1-(4/3){\bf S}_1\cdot{\bf S}_1$. Collecting all the ${\bf S}_1\cdot{\bf S}_1$ interaction terms resulting from the pair-hopping, Hund's coupling, and density interaction terms corresponding to $J_{\rm H}$, as obtained in Eqs. \ref{phterm_2}, \ref{jhterm_2}, \ref{jhterm_3}, yields an exact cancellation. 
%-\frac{J_{\rm H}}{2} \sum_{\mu<\nu} c_{\mu 1}^2 c_{\nu 1}^2 [(n_{1\uparrow} + n_{1\downarrow}) + 2 n_{1\uparrow} n_{1\downarrow}] = \frac{2J_{\rm H}}{9} {\bf S}_1. {\bf S}_1 - \frac{J_{\rm H}}{3} n_1
% -\frac{J_{\rm H}}{2} \frac{1}{3} \left [n_1 + \left (n_1 - \frac{4}{3} {\bf S}_1. {\bf S}_1 \right )\right ]  

Similarly, considering the other two intra-pseudo-orbital cases ($l=m=l'=m'=2,3$) in the $J=3/2$ sector, we obtain from the three explicitly $J_{\rm H}$ interaction terms:
\begin{eqnarray}
\frac{J_{\rm H}}{2}\sum_{\mu\ne\nu} (\psi_\mu^\dagger \psi_\nu)^2 &\leadsto & 
-\frac{4J_{\rm H}}{9} {\bf S}_2\cdot{\bf S}_2 + \frac{J_{\rm H}}{3} S_{2z} S_{2z} - J_{\rm H} S_{3z} S_{3z} \nonumber \\ 
-2J_{\rm H} \sum_{\mu<\nu} {\bf S}_\mu\cdot{\bf S}_\nu &\leadsto &  
\frac{J_{\rm H}}{18} {\bf S}_2\cdot{\bf S}_2 + \frac{J_{\rm H}}{3} S_{2z} S_{2z} +
\frac{J_{\rm H}}{2} {\bf S}_3\cdot{\bf S}_3 - J_{\rm H} S_{3z} S_{3z} \nonumber \\ 
-\frac{J_{\rm H}}{2} \sum_{\mu<\nu} n_{\mu} n_{\nu} &\leadsto & 
\frac{J_{\rm H}}{6} {\bf S}_2\cdot{\bf S}_2 + \frac{J_{\rm H}}{6} {\bf S}_3.{\bf S}_3 
- \frac{J_{\rm H}}{4} (n_2 + n_3) 
\label{jh_3/2}
\end{eqnarray}
explicitly showing the symmetry breaking contributions $(S_{lz} S_{lz})$ from the pair-hopping and Hund's coupling interaction terms, as expected from the SU(2) transformation analysis in Sec. III, in the absence of the special symmetry in the $J=3/2$ sector. However, since $S_\alpha S_\alpha = (1/4)[(n_\uparrow + n_\downarrow) - 2 n_\uparrow n_\downarrow]=(1/3){\bf S}\cdot{\bf S}$ for all three components $\alpha=x,y,z$ of $S=1/2$ quantum spin operators $S_\alpha =(1/2)\psi^\dagger \sigma_\alpha \psi$, there is no true magnetic anisotropy even if classically anisotropic terms such as $S_{lz}S_{lz}$ are present, as in Eq. (\ref{jh_3/2}). 

%Such classically anisotropic interaction terms are also generated when the octahedral cubic symmetry is lifted by the tetragonal distortion in $\rm Sr_2IrO_4$.\cite{iridate_four} 

Substituting $S_{lz} S_{lz} = (1/3){\bf S}_l\cdot{\bf S}_l$ in Eq. (\ref{jh_3/2}) yields an exact cancellation of the three explicitly $J_{\rm H}$ interaction terms in the $J=3/2$ sector also. Therefore, the Hubbard like (or equivalently ${\bf S}_l\cdot{\bf S}_l$) intra-pseudo-orbital interaction terms in all three sectors result only from the $U$ and $U'$ terms in Eq. (\ref{inter}). Using similar analysis as above, and dropping one-particle density terms as in Eq. (\ref{jhterm_3}), one obtains (for site $i$):
\begin{eqnarray}
U\sum_{\mu}{n_{\mu\uparrow}n_{\mu\downarrow}} + U^\prime \sum_{\mu < \nu} n_\mu n_\nu 
& \leadsto & \left (\frac{U+2U'}{3} \right ) n_{1\uparrow} n_{1\downarrow} + \left (\frac{U+U'}{2} \right ) (n_{2\uparrow}n_{2\downarrow} + n_{3\uparrow}n_{3\downarrow}) 
%& - & \left (\frac{U-U'}{3} \right ) n_1 - \left (\frac{U-U'}{4} \right ) (n_2 + n_3) .
\end{eqnarray}

From the above discussion it follows that true magnetic anisotropy results only from the inter-orbital anisotropic interaction terms such as $S_{lz}S_{l'z}$ with $l'\ne l$. Including inter-pseudo-orbital cases such as $l$=$m$, $l'$=$m'$ and $l$=$m'$, $l'$=$m$ in Eqs. (\ref{phterm_2}) and (\ref{jhterm_2}), and keeping all interaction terms relevant for the present study (Hubbard, Hund's coupling, and density), we obtain (for site $i$):
\begin{eqnarray}
\mathcal{H}_{\rm int} (i) & = & \left ( \frac{U + 2 U^\prime}{3} \right ) n_{1\uparrow} n_{1\downarrow} + \left ( \frac{U +  U^\prime}{2} \right ) \left [n_{2\uparrow} n_{2\downarrow} + n_{3\uparrow} n_{3 \downarrow} \right ] \nonumber \\
&-& \left ( \frac{U-U^\prime}{3} \right ) 2 {\bf S}_1 \cdot {\bf S}_2 
- \left (\frac{U-U^\prime - 2J_{\rm H}}{3} \right ) \big [(2{\bf S}_1 + {\bf S}_2) \cdot {\bf S}_3 \big ] + 2J_{\rm H}[S_1 ^z S_2 ^z - S_1 ^z S_3 ^z] \nonumber \\
& + & \left (\frac{U + 5U^\prime -3J_{\rm H}}{6} \right ) [n_1 n_2 + n_1 n_3] + \left (\frac{U + 11U^\prime -6J_{\rm H}}{12} \right ) n_2 n_3 
\label{h_int_detail}
\end{eqnarray}
where ${\bf S}_m = \psi_m ^\dagger \frac{\mbox{\boldmath $\tau$}}{2} \psi_m$ and $n_m = \psi_m ^\dagger {\bf 1} \psi_m = n_{m \uparrow} + n_{m \downarrow}$ are the spin and charge density operators. Using the spherical symmetry condition ($U^\prime$=$U$-$2J_{\mathrm H}$), the transformed interaction Hamiltonian (\ref{h_int_detail}) simplifies to:
\begin{eqnarray}
{\mathcal H}_{\rm int}(i) &=& \left( U - \frac{4}{3} J_{\rm H} \right) n_{1 \uparrow} n_{1 \downarrow} + \left( U - J_{\rm H} \right) \left[ n_{2 \uparrow}  n_{2 \downarrow} +  n_{3 \uparrow} n_{3 \downarrow} \right] \nonumber \\
&-& \frac{4}{3} J_{\rm H} {\bf S}_1\cdot{\bf S}_2 + 2 J_{\rm H} \left [ \mathcal{S}_{1}^z \mathcal{S}_{2}^z - \mathcal{S}_{1}^z \mathcal{S}_{3}^z \right] \nonumber \\
&+& \left( U-\frac{13}{6} J_{\rm H} \right) \left[ n_{1} n_{2} + n_{1} n_{3} \right] + 
\left( U-\frac{7}{3} J_{\rm H} \right) n_{2} n_{3} .
\label{h_int}
\end{eqnarray}
%The symmetry features of the interaction terms above are consistent with a general pseudo-spin rotation symmetry analysis which shows that the Hund's coupling ($J_{\rm H}$) and pair-hopping ($J_{\rm H}$) interaction terms in Eq. (\ref{inter}) explicitly break this symmetry systematically, while the Hubbard $(U)$ and density $(U')$ interaction terms do not.\cite{iridate_four} 

In the above equation, the Hubbard-like terms ${\cal U}_m n_{m\uparrow} n_{m\downarrow} \sim -{\cal U}_m {\bf S}_m\cdot{\bf S}_m$, the Hund's-coupling-like term ${\bf S}_1\cdot{\bf S}_2$, and the density terms $n_l n_m$, are all invariant under pseudo-spin rotation and the corresponding SU(2) transformation $\psi_m \rightarrow \psi_m ' = [U]\psi_m$. Therefore, only the interaction terms $S_1 ^z S_2 ^z$ and $S_1 ^z S_3 ^z$ between moments in the $J$=1/2 and $J$=3/2 sectors are responsible for the magneto-crystalline anisotropy in $\rm Sr_2 Ir O_4$, highlighting the importance of the weak magnetism in the nominally filled $J$=3/2 sector due to the mixing between the two sectors. Magnetic anisotropy will not survive in the large SOC limit when the two sectors become effectively decoupled. As shown below, an easy $x$-$y$ plane anisotropy is obtained from the dominant term ($-2J_{\rm H} S_1 ^z S_3 ^z$) due to opposite sign of $J$=3/2 sector moment.

\section{Easy x-y plane anisotropy}
We consider the various interaction terms in Eq. (\ref{h_int}) in the Hartree-Fock (HF) approximation, focussing on the staggered field terms corresponding to ($\pi,\pi$) ordered AF state on the square lattice. For general ordering direction with components {\boldmath $\Delta_l$}= $(\Delta_l ^x,\Delta_l ^y,\Delta_l ^z)$, the staggered field term for sector $l$ in the pseudo-orbital basis is given by:  
% , focussing on the 
\begin{equation}
\mathcal{H}_{\rm sf} (l) = 
\sum_{{\bf k} s}  \psi_{{\bf k}ls}^{\dagger} 
\begin{pmatrix} -s \makebox{\boldmath $\tau\cdot\Delta_l$}
\end{pmatrix} \psi_{{\bf k}ls} 
= \sum_{{\bf k} s} -s \psi_{{\bf k}ls}^{\dagger} 
\begin{pmatrix} \Delta_l ^z & \Delta_l ^x - i \Delta_l ^y \\
\Delta_l ^x + i \Delta_l ^y & -\Delta_l ^z  \\
\end{pmatrix} \psi_{{\bf k}ls}
\label{gen_ord_dirn} 
\end{equation}  
where $\psi_{{\bf k}ls} ^\dagger = (a_{{\bf k}ls\uparrow} ^\dagger \; \; a_{{\bf k}ls\downarrow} ^\dagger)$, $s=\pm 1$ for the two sublattices A/B, and the staggered field components $\Delta_{l=1,2,3} ^{\alpha=x,y,z}$ are self-consistently determined from:
\begin{eqnarray}
2 \Delta_1 ^\alpha &=& {\mathcal U}_1 m_1 ^\alpha + \frac{2J_{\rm H}}{3} m_2 ^\alpha + J_{\rm H} (m_3 ^\alpha - m_2 ^\alpha) \delta_{\alpha z} \nonumber \\
2 \Delta_2 ^\alpha &=& {\mathcal U}_2 m_2 ^\alpha + \frac{2J_{\rm H}}{3} m_1 ^\alpha - J_{\rm H} m_1 ^\alpha \delta_{\alpha z} \nonumber \\
2 \Delta_3 ^\alpha &=& {\mathcal U}_3 m_3 ^\alpha + J_{\rm H} m_1 ^\alpha \delta_{\alpha z}  
\label{selfcon}
\end{eqnarray}
in terms of the staggered pseudo-spin magnetization components $m_{l=1,2,3} ^{\alpha=x,y,z}$. In practice, it is easier to choose set of ${\bf \Delta}_{l=1,2,3}$ and self-consistently determine the Hubbard-like interaction strengths ${\mathcal U}_{l=1,2,3}$ such that ${\mathcal U}_1 = U - \frac{4}{3} J_{\rm H}$ and ${\mathcal U}_2 = {\mathcal U}_3 =  U - J_{\rm H}$ using Eq. (\ref{selfcon}). The interaction strengths are related by ${\mathcal U}_{l=2,3} = {\mathcal U}_{l=1} + J_{\rm H}/3$.

Transforming the staggered-field term back to the three-orbital basis $(yz\sigma,xz\sigma,xy\bar{\sigma})$, and including the crystal field, SOC and band terms,\cite{watanabe_PRL_2010} we have considered $\mathcal {H}_{\rm HF} = \mathcal{H}_{\rm SO} + \mathcal{H}_{\rm band+cf} + \mathcal{H}_{\rm sf}$ in our band structure and spin fluctuation analysis, where
\begin{eqnarray}
\mathcal{H}_{\rm band+cf} &=& 
\sum_{{\bf k} \sigma s} \psi_{{\bf k} \sigma s}^{\dagger} \left [ \begin{pmatrix}
{\epsilon_{\bf k} ^{yz}}^\prime & 0 & 0 \\
0 & {\epsilon_{\bf k} ^{xz}}^\prime & 0 \\
0 & 0 & {\epsilon_{\bf k} ^{xy}}^\prime + \epsilon_{xy} \end{pmatrix} \delta_{s s^\prime}
+ 
\begin{pmatrix}
\epsilon_{\bf k} ^{yz} & \epsilon_{\bf k} ^{yz|xz} & 0 \\
-\epsilon_{\bf k} ^{yz|xz} & \epsilon_{\bf k} ^{xz} & 0 \\
0 & 0 & \epsilon_{\bf k} ^{xy} \end{pmatrix} \delta_{\bar{s} s^\prime } \right]
\psi_{{\bf k} \sigma s^\prime} 
\label{three_orb_two_sub}
\end{eqnarray} 
in the composite three-orbital, two-sublattice basis. The crystal field induced tetragonal splitting is included as the $xy$ orbital energy offset $\epsilon_{xy}$ from the degenerate $yz/xz$ orbitals. The different hopping terms in Eq. (\ref{three_orb_two_sub}) connecting the same ($s=s'$) and opposite ($s\ne s'$) sublattice(s) are given by: 
\begin{eqnarray}
\epsilon_{\bf k} ^{xy} &=& -2t_1(\cos{k_x} + \cos{k_y}) \nonumber \\
{\epsilon_{\bf k} ^{xy}} ^{\prime} &=& - 4t_2\cos{k_x}\cos{k_y} - \> 2t_3(\cos{2{k_x}} + \cos{2{k_y}}) \nonumber \\
\epsilon_{\bf k} ^{yz} &=& -2t_5\cos{k_x} -2t_4 \cos{k_y} \nonumber \\
\epsilon_{\bf k} ^{xz} &=& -2t_4\cos{k_x} -2t_5 \cos{k_y}  \nonumber \\
\epsilon_{\bf k} ^{yz|xz} &=&  -2t_{m}(\cos{k_x} + \cos{k_y}) . 
\end{eqnarray}
Here $t_1$, $t_2$, $t_3$ are respectively the first, second, and third neighbor hopping terms for the $xy$ orbital. For the $yz$ ($xz$) orbital, $t_4$ and $t_5$ are the NN hopping terms in $y$ $(x)$ and $x$ $(y)$ directions, respectively. Mixing between $xz$ and $yz$ orbitals is represented by the NN hopping term $t_m$. We have taken values of the tight-binding parameters ($t_1$, $t_2$, $t_3$, $t_4$, $t_5$, $t_{\rm m}$, $\epsilon_{xy}$, $\lambda$) = (1.0, 0.5, 0.25, 1.028, 0.167, 0.2, -0.7, 1.35) in units of $t_1$, where the energy scale $t_1$ = 280 meV. Using above parameters, the calculated electronic band structure shows AFM insulating state and mixing between pseudo-orbital sectors.\cite{watanabe_PRL_2010,iridate_one}

For the band dispersion terms in Eq. (\ref{three_orb_two_sub}) and henceforth, we have used the crystal (global) coordinate axes referred to as $x,y,z$ for convenience. The spin coordinate axes are chosen to align with the crystal axes. If required, a site-dependent spin rotation allows one to transform back to the local spin coordinate axes common with the octahedral axes due to SOC.

\subsection*{Canted AFM state}
\begin{figure}
\vspace*{0mm}
\hspace*{0mm}
\psfig{figure=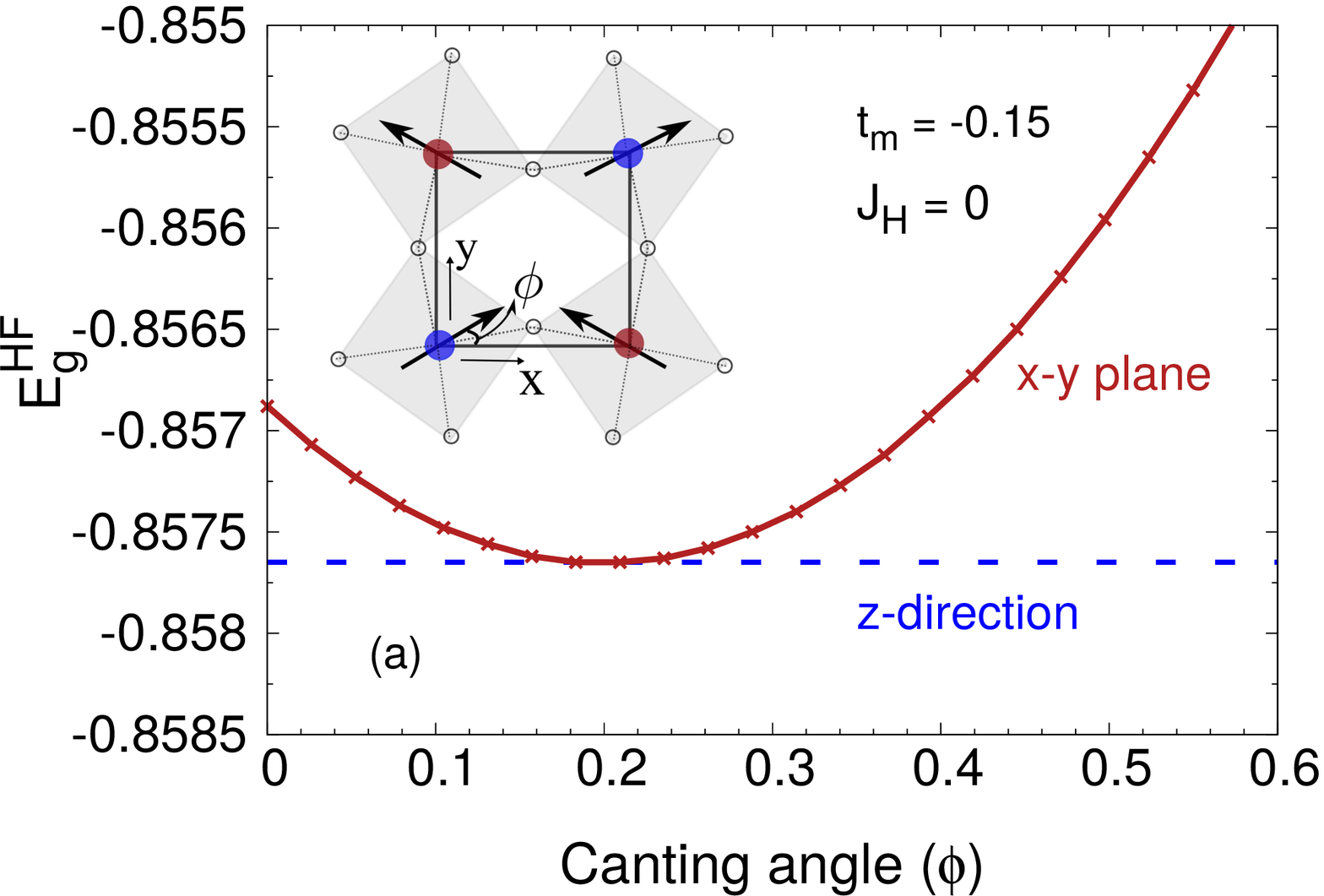,angle=0,width=80mm} 
\psfig{figure=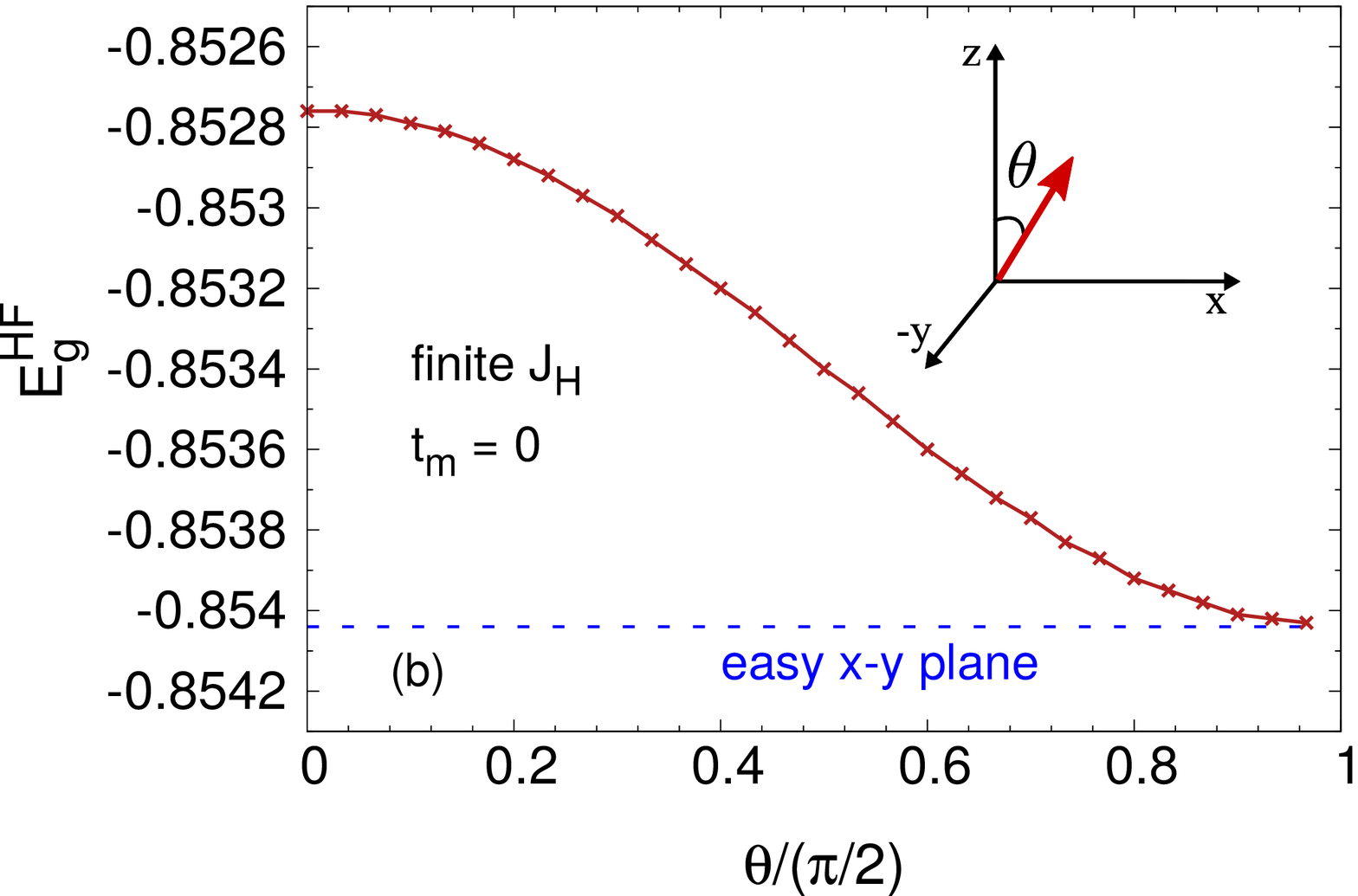,angle=0,width=80mm} 
\vspace{-0mm}
\caption{Variation of AFM state energy per state (in units of $t_1$) (a) for $x$-$y$ plane ordering with canting angle $\phi$ including finite $yz-xz$ orbital mixing hopping showing degeneracy at the optimal canting angle with the $z$-ordered AFM state, and (b) with the staggered field polar angle $\theta$ showing easy $x$-$y$ plane anisotropy for finite Hund's coupling.} 
\label{gse}
\end{figure}

The octahedral-rotation-induced orbital mixing hopping term ($t_{\rm m}$) between $yz$ and $xz$ orbitals generates PD ($S_i^z S_j^z$) and DM [${\hat z}\cdot({\bf S}_i \times {\bf S}_j)$] anisotropic interactions in the strong coupling limit.\cite{iridate_one} However, the AFM-state energy is invariant with respect to change of ordering direction from $z$ axis to $x$-$y$ plane provided spins are canted at the optimal canting angle, thus preserving the gapless Goldstone mode. Fig. \ref{gse}(a) shows the variation of AFM-state energy with canting angle ($\phi$) for ordering in the $x$-$y$ plane. The energy minimum at the optimal canting angle is exactly degenerate with the energy for $z$-direction ordering. This absence of true anisotropy in $\rm Sr_2 Ir O_4$ due to octahedral rotation alone is consistent with the general gauge transformation analysis showing that the spin-dependent hopping terms arising from the orbital-mixing terms in the three-orbital model can be gauged away.

%\cite{iridate_four} 

\subsection*{$J_{\rm H}$-induced easy-plane anisotropy}
The Hund's-coupling-induced easy-plane magnetic anisotropy is explicitly shown in Fig.\ref{gse}(b) by the variation of AFM-state energy with polar angle $\theta$ corresponding to staggered field orientation in the $x$-$z$ plane, with $\Delta_1 ^z = (\Delta +\Delta_{\rm ani})\cos \theta$ and $\Delta_1 ^x = \Delta \sin \theta$. Here $\Delta$ represents the spin-rotationally-symmetric part $({\mathcal U}_1 m_1 + \frac{2J_{\rm H}}{3} m_2)/2$ of the staggered field and $\Delta_{\rm ani} = J_{\rm H}(m_3^z - m_2^z)/2$ is the symmetry-breaking term, as seen from Eq. (\ref{selfcon}). We have taken $\Delta$=0.9, $\Delta_{\rm ani}$=$-0.01$, and the orbital mixing hopping term $t_{\rm m}$ has been set to zero for simplicity. Converting from energy per state as shown in Fig.\ref{gse} to simply energy per site ($\times$10 occupied states for each ${\bf k}$), yields the magnetic anisotropy energy $E_{\rm g}^{\rm HF}(z)-E_{\rm g}^{\rm HF}(x)\approx 0.012$, which is comparable to the $\Delta_{\rm ani}$ magnitude. The simplified analysis presented in this section, with staggered field only for the $l$=1 orbital, serves to explicitly illustrate the magnetic anisotropy features within our three-band-model calculation. 

% The $\Delta_{\rm ani}$ value taken here is realistic, as shown in the next section where the magnetization values are reported. 

\section{Magnon excitations and anisotropy gap}
In view of the Hund's-coupling-induced easy $x$-$y$ plane anisotropy as discussed above, we consider the $x$-ordered AFM state. The magnon propagator corresponding to transverse spin fluctuations should therefore yield one gapless mode ($y$ direction) and one gapped mode ($z$ direction). Accordingly, we consider the time-ordered magnon propagator: 
\begin{equation}
\chi ({\bf q},\omega) = \int dt \sum_{i} e^{i\omega(t-t^\prime)}
e^{-i{\bf q}.({\bf r}_i - {\bf r}_j)}  
\langle \Psi_0 | T [ S_{i m} ^{\alpha} (t) S_{j n} ^{\beta} (t^\prime) ] | \Psi_0 \rangle
\label{chi}
\end{equation}
involving the transverse $\alpha,\beta=y,z$ components of the pseudo-spin operators $S_{im}^{\alpha}$ and $S_{jn}^{\beta}$ for pseudo orbitals $m$ and $n$ at lattice sites $i$ and $j$. 

In the random phase approximation (RPA), the magnon propagator is obtained as:
\begin{equation}
[\chi({\bf q},\omega)] = \frac{[\chi^0({\bf q},\omega)]}
{1 - 2 [\mathcal{U}] [\chi^0({\bf q},\omega)]}
\label{eq:spin_prop}  
\end{equation}
where the bare particle-hole propagator:
\begin{equation}
[\chi^0 ({\bf q},\omega)]_{a b} ^{\alpha \beta} = \frac{1}{4} \sum_{{\bf k}} \left [ 
\frac{ 
\langle \varphi_{\bf k-q} | \tau^\alpha | \varphi_{\bf k} \rangle_a
\langle \varphi_{\bf k} | \tau^\beta | \varphi_{\bf k-q} \rangle_b 
} 
{E^+_{\bf k-q} - E^-_{\bf k} + \omega - i \eta }
+ \frac{
\langle \varphi_{\bf k-q} | \tau^\alpha | \varphi_{\bf k} \rangle_a
\langle \varphi_{\bf k} | \tau^\beta | \varphi_{\bf k-q} \rangle_b 
} 
{E^+_{\bf k} - E^-_{\bf k-q} - \omega - i \eta } \right ]
\end{equation}
was evaluated in the composite spin-orbital-sublattice basis (2 spin components $\alpha,\beta=y,z$ $\otimes$ 3 pseudo orbitals $m=1,2,3$ $\otimes$ 2 sublattices $s,s'=$ A,B) by integrating out the fermions in the $(\pi,\pi)$ ordered state. Here $E_{\bf k}$ and $\varphi_{\bf k}$ are the eigenvalues and eigenvectors of the Hamiltonian matrix in the pseudo-orbital basis, the indices $a,b=1,6$ correspond to the orbital-sublattice subspace, and the superscript $+(-)$ refers to particle (hole) energies above (below) the Fermi energy. The amplitudes $\varphi^m _{{\bf k}\tau}$ were obtained by projecting the ${\bf k}$ states in the three-orbital basis on to the pseudo-orbital basis states $|m,\tau = \uparrow,\downarrow \rangle$  corresponding to the $J=1/2$ and $3/2$ sector states, as given below:
\begin{eqnarray}
\varphi_{{\bf k}\uparrow}^1 = \frac{1}{\sqrt{3}} \left( \phi^{yz}_{{\bf k}\downarrow} - i\phi^{xz}_{{\bf k}\downarrow} + \phi^{xy}_{{\bf k}\uparrow}\right) \;\;\;\;\;\; & & 
\varphi_{{\bf k}\downarrow}^1 = \frac{1}{\sqrt{3}} \left( \phi^{yz}_{{\bf k}\uparrow} + i \phi^{xz}_{{\bf k}\uparrow} - \phi^{xy}_{{\bf k}\downarrow}\right) \nonumber \\ 
\varphi_{{\bf k}\uparrow}^2 = \frac{1}{\sqrt{6}} \left( \phi^{yz}_{{\bf k}\downarrow} - i\phi^{xz}_{{\bf k}\downarrow} - 2 \phi^{xy}_{{\bf k}\uparrow}\right)  \;\;\;\; & & 
\varphi_{{\bf k}\downarrow}^2 = \frac{1}{\sqrt{6}} \left( \phi^{yz}_{{\bf k}\uparrow} + i \phi^{xz}_{{\bf k}\uparrow} + 2 \phi^{xy}_{{\bf k}\downarrow}\right) \nonumber \\
\varphi_{{\bf k}\uparrow}^3 = \frac{1}{\sqrt{2}} \left( \phi^{yz}_{{\bf k}\downarrow} + i\phi^{xz}_{{\bf k}\downarrow} \right) \;\;\;\;\;\;\;\;\;\;\;\;\;\;\; & & 
\varphi_{{\bf k}\downarrow}^3 = \frac{1}{\sqrt{2}} \left( \phi^{yz}_{{\bf k}\uparrow} - i \phi^{xz}_{{\bf k}\uparrow} \right)
\label{proj_ampl}
\end{eqnarray}
in terms of the amplitudes $\phi^{\mu}_{{\bf k}\sigma}$ in the three-orbital basis $(\mu = yz,xz,xy)$. 

The rotationally invariant Hubbard- and Hund's coupling-like terms having the form $S_{im}^\alpha S_{in} ^\beta \delta_{\alpha\beta}$ are diagonal in spin components ($\alpha = \beta$). The on-site Coulomb interaction terms are also diagonal in the sublattice basis ($s=s'$). The interaction matrix $[\mathcal{U}]$ in Eq. (\ref{eq:spin_prop}) is therefore obtained as:
\begin{equation}
[\mathcal{U}] = \begin{pmatrix} \mathcal{U}_1 & \frac{2}{3} J_{\rm H} & 0 \\
\frac{2}{3} J_{\rm H} & \mathcal{U}_2 & 0 \\ 0 & 0 & \mathcal{U}_3 
\end{pmatrix} \delta_{\alpha\beta} \delta_{ss'} + 
\begin{pmatrix} 0 & -J_{\rm H} & J_{\rm H}  \\
-J_{\rm H} & 0 & 0 \\ J_{\rm H} & 0 & 0 
\end{pmatrix} \delta_{\alpha z} \delta_{\beta z} \delta_{ss'} 
\end{equation}
in the pseudo-orbital basis. While the first interaction term above preserves spin rotation symmetry, the second interaction term (corresponding to the $S_{im}^z S_{in}^z$ terms in Eq. \ref{h_int}) breaks rotation symmetry and is responsible for easy $x$-$y$ plane anisotropy. The spin wave energies are calculated from the pole condition $1-2\xi_{\bf q}(\omega)=0$ of Eq. \ref{eq:spin_prop}, where $\xi_{\bf q}(\omega)$ are the eigenvalues of the $[{\cal U}] [\chi^0 ({\bf q},\omega)]$ matrix. The $12 \times 12$ $[\chi^0 ({\bf q},\omega)]$ matrix was evaluated by performing the $\bf k$ sum over the 2D Brillouin zone divided into a 300 $\times$ 300 mesh.

\begin{figure}
\vspace*{0mm}
\hspace*{0mm}
\psfig{figure=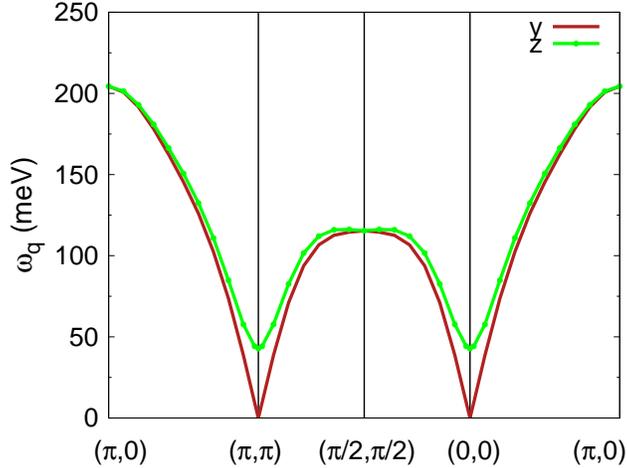,angle=0,width=90mm}
\vspace{0mm}
\caption{The calculated magnon dispersion in the three-orbital model with staggered field in the $x$ direction. The easy $x$-$y$ plane anisotropy arising from Hund's coupling results in one gapless mode and one gapped mode corresponding to transverse fluctuations in the $y$ and $z$ directions, respectively.}
\label{sw}
\end{figure}

%\section{magnon Dispersion}
The calculated magnon energies in the $x$-ordered AFM state are shown in Fig. \ref{sw}. Here we have taken staggered field values $\Delta_{l=1,2,3}^x= (0.92,0.08,-0.06)$ in units of $t_1$, which ensures self-consistency for all three orbitals, with the given relations ${\mathcal U}_2$=${\mathcal U}_3$=${\mathcal U}_1$+$J_{\rm H}/3$. Using the calculated sublattice magnetization values $m_{l=1,2,3}^x$=(0.65,0.005,-0.038), we obtain ${\mathcal U}_{l=1,2,3}$=(0.80,0.83,0.83) eV, which finally yields $U$=$\mathcal U_1$+$\frac{4}{3}J_{\rm H}$=0.93 eV for $J_{\rm H}$=0.1 eV. 

The magnon dispersion clearly shows the Goldstone mode and the gapped mode, corresponding to transverse spin fluctuations in the $y$ and $z$ directions, respectively. The easy $x$-$y$ plane anisotropy arising from Hund's coupling results in energy gap $\approx 40$ meV for the out-of-plane ($z$) mode. The two modes are degenerate at $(\pi,0)$ and $(\pi/2,\pi/2)$. The excitation energy at $(\pi,0)$ is approximately twice that at $(\pi/2,\pi/2)$, and the strong zone-boundary dispersion in this iridate compound was ascribed to finite-$U$ and finite-SOC effects.\cite{iridate_one} The calculated magnon dispersion and energy gap are in very good agreement with RIXS measurements.\cite{kim1_PRL_2012,kim_NATCOMM_2014,pincini_PRB_2017,porras_PRB_2019}

%Without $J_{\rm H}$, both modes are found to be gapless with degenerate energies at all points.

The electron fillings in the different pseudo orbitals are obtained as $n_{l=1,2,3} \approx (1.064,1.99,1.946)$. Finite mixing between the $J$=1/2 and 3/2 sectors is reflected in the small deviations from ideal fillings (1,2,2) and also in the very small magnetic moment values for $l=2,3$ as given above, which play a crucial role in the expression of magnetic anisotropy and magnon gap in view of the anisotropic $J_{\rm H}$ interaction terms in Eq. (\ref{h_int}). The values $\lambda$=0.38 eV, $U$=0.93 eV, and $J_{\rm H}$=0.1 eV taken above lie well within the estimated parameter range for $\rm Sr_2 Ir O_4$.\cite{igarashi_PRB_2014,zhou_PRX_2017}

\begin{figure}
\vspace*{0mm}
\hspace*{0mm}
\psfig{figure=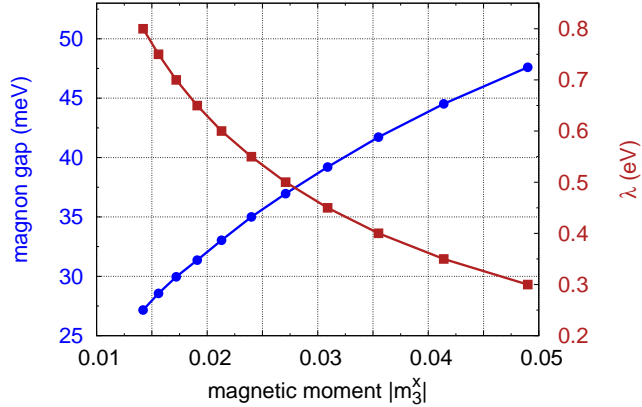,angle=0,width=90mm}
\vspace{0mm}
\caption{Variation of magnon gap with magnetic moment $|m_3^x|$ in the $l$=3 orbital (blue curve). The magnitude of $|m_3^x|$ decreases with SOC strength (red curve) due to suppression of mixing between $J$=1/2 and 3/2 sectors. Here $(U,J_{\rm H})$=(0.93 eV, 0.1 eV).}
\label{fig3b}
\end{figure}

We have investigated the crucial role of the small $J$=3/2 sector magnetic moment on the magnon gap by studying the variation with SOC strength which effectively controls the mixing between $J$=1/2 and 3/2 sectors. Fig. \ref{fig3b} shows that the magnon gap sharply increases with the dominant magnetic moment $|m_3^x|$, highlighting the finite-SOC effect on the experimentally observed out-of-plane magnon gap in $\rm Sr_2 Ir O_4$. The opposite sign of the magnetic moment $m_3^x$ as compared to $m_1^x$ (due to spin-orbital entanglement) plays a vital role in the easy-plane anisotropy. It should be noted that the tetragonal splitting $\epsilon_{xy}$ weakly affects the magnon gap through the $J$=3/2 sector magnetic moments. 

The magnon gap in Fig. \ref{sw} is related to the anisotropy term $\Delta_{\rm ani}$ (see discussion of Fig. \ref{gse}), and we consider here an analytical expression relating the two which can be derived for a simple model. For the anisotropic Hubbard model,\cite{chatterji_PRB_2007} using the spin fluctuation analysis for the AFM state on the square lattice,\cite{singh_PRB_1990} we obtain: 
\begin{equation}
\frac{\omega_{\rm gap}}{\omega_{\rm max}} = \sqrt {\left (\frac{U_{\rm ani}}{t} \right )
\left (\frac{U}{4t} \right ) }
\end{equation}
in the strong coupling limit $(U\gg t$), where $\omega_{\rm max}=2(4t^2/U)$, and $U_{\rm ani} \approx 2\Delta_{\rm ani}$ is the anisotropy term included in the Hubbard model. With $\Delta_{\rm ani}/t=0.01$ (as in Fig. \ref{gse}(b) discussion) and $U/4t=1$ for simplicity, the ratio is obtained as $\approx 1/7$, which is in good agreement with Fig. \ref{sw}. It should be noted that a magnon gap of $\sim 4$ meV (one-tenth of that in Fig. \ref{sw}) would correspond to an anisotropy term $\Delta_{\rm ani}/t \sim 10^{-4}$ (one-hundredth of above), which is about 0.03 meV using the hopping energy scale $t_1=280$ meV.

%Here the interaction term considered is $(U+U_{\rm ani})$ for the local moment ($\langle S_i ^z \rangle$) and $U$ in the ladder sum for the transverse ($S_i ^x, S_i ^y$) spin fluctuation propagator.

\section{Comparison with pure orbital based approach}
A fully self-consistent approach within the pure ($t_{\rm 2g}$) orbital basis, wherein all orbital off-diagonal spin and charge condensates are included (along with the diagonal condensates) in the Coulomb interaction contributions, has been applied recently to the $\rm Ca_2 RuO_4$ compound with electron filling $n=4$ to investigate the complex interplay due to intimately intertwined roles of SOC, Coulomb interactions, and structural distortions.\cite{ruthenate_one} We discuss below the relevant results obtained by extending this approach to the $\rm Sr_2 IrO_4$ compound with $n=5$.

\begin{table}[t] \vspace*{-10mm}
\caption{Self consistently determined magnetization and density values for the three pure ($\mu=yz,xz,xy$) and three pseudo ($l=1,2,3$) orbitals on the two sublattices ($s$=A/B), showing the simplified structure in the pseudo orbital basis. The parameter set is same as in previous section.} 
\centering % centering table
\begin{tabular}{l c c c c} \\  % creating 5 columns 
\hline\hline % inserting double-line
$\mu$ (s) & $m_\mu^x$ & $m_\mu^y$ & $m_\mu^z$ & $n_\mu$ \\ [0.5ex]
\hline % inserts single-line
$yz$ (A/B) & $\pm$0.24 & 0.0 & 0.0 & 1.62 \\[1ex]
$xz$ (A/B) & $\mp$0.24 & 0.0 & 0.0 & 1.62 \\[1ex]
$xy$ (A/B) & $\mp$0.18 & 0.0 & 0.0 & 1.75 \\[1ex] \hline
\end{tabular} \hspace{10mm}
\begin{tabular}{l c c c c} \\  % creating 5 columns 
\hline\hline % inserting double-line
$l$ (s) & $m_l^x$ & $m_l^y$ & $m_l^z$ & $n_l$ \\ [0.5ex]
\hline % inserts single-line
$1$ (A/B) & $\pm$0.69 & 0.0 & 0.0 & 1.04 \\[1ex]
$2$ (A/B) & $\pm$0.006 & 0.0 & 0.0 & 1.99 \\[1ex]
$3$ (A/B) & $\mp$0.023 & 0.0 & 0.0 & 1.97 \\[1ex] \hline
\end{tabular}\label{table1}
\end{table}

Most importantly, we have explicitly confirmed the $J_{\rm H}$ induced easy-plane anisotropy. The anisotropy is completely absent when $J_{\rm H}$=0 even for finite $\epsilon_{xy}$, and the fully self consistent calculation (with octahedral rotation turned off for simplicity) yields degenerate solutions corresponding to arbitrary orientation of the ($l$=1) pseudo-spin moment, confirming that finite $\epsilon_{xy}$ is not the source of any anisotropy. Moreover, we do not find any indication of easy-axis anisotropy within the basal plane even in the combined presence of Hund's coupling ($J_{\rm H}$), octahedral rotation ($t_m$), and tetragonal distortion ($\epsilon_{xy}$). When a small octahedral tilting about crystal $a$ axis is included, we obtain easy-axis anisotropy along crystal $b$ axis and pseudo-spin canting in the $z$ direction, which can be understood in terms of the induced DM interaction resulting from the interplay between octahedral tilting and SOC.\cite{ruthenate_one}

%as inferred from degenerate self consistent solutions obtained  arbitrary basal plane orientation of the $l=1$ pseudo-spin moment.
% with no indication of easy-axis anisotropy when the . 

The simplicity in the pseudo orbital basis is illustrated by comparing the magnetization and density values for the pure and pseudo orbitals. The projected amplitudes (Eq. \ref{proj_ampl}) were used to convert from pure to pseudo orbitals. The results are presented in Table I, showing the much simpler structure in the pseudo orbital basis, with only $l$=1 sector magnetically active and nearly nominal density values. Furthermore, while the orbital off-diagonal condensates in the pure orbital basis are finite and yield the Coulomb SOC renormalization, all off-diagonal condensates in the pseudo orbital basis were found to be negligible, which is expected as the SOC term is diagonal. The Coulomb renormalized SOC gap in this basis is directly obtained from the orbital diagonal terms.\cite{mohapatra_JMMM_2020} The reduced $m_l^x$ value for $l$=3 in Table I compared with the previous section (see Fig. \ref{fig3b}) is consistent with the Coulomb enhanced SOC value (from $\sim 0.4$ to 0.6 eV) in the fully self consistent calculation.

\section{Discussion}
Finally, a critical comparison with other approaches for studying magnetic anisotropy in $\rm Sr_2IrO_4$ is presented below. In Ref. [10], the $\rm IrO_6$ octahedral rotation induced PD $(J_z S_i^z S_j^z)$ and DM $[{\bf D} \cdot ({\bf S}_i \times {\bf S}_j)]$ terms were considered as the dominant anisotropic interactions, which were shown to be gauged out by a  site-dependent rotation of the spin operators (${\bf S} \rightarrow {\bf {\tilde S}}$), resulting in no true magnetic anisotropy in the absence of Hund's coupling. After including the $J_{\rm H}$ induced corrections, the resulting anisotropic interaction terms were obtained as:
\begin{equation}
\mathcal{\tilde H_{\rm ani}} = -\Gamma_1 {\tilde S}_i^z {\tilde S}_j^z \pm \Gamma_2 ({\tilde S}_i^x {\tilde S}_j^x - {\tilde S}_i^y {\tilde S}_j^y)
\end{equation}
in terms of the rotated spin operator ${\bf {\tilde S}}$, where the $\pm$ sign corresponds to bond along the $x(y)$ direction. Presence of tetragonal distortion was found to affect the coefficient $\Gamma_1$. 

Together with the Heisenberg AFM interaction, the first term with $\Gamma_1 >0$ leads to easy $a-b$ plane magnetic ordering and magnon gap for out-of-plane fluctuations, whereas the much smaller easy-axis anisotropy and magnon gap for in-plane fluctuations were ascribed to the second term ($\Gamma_2 \ll \Gamma_1$). In Ref. [17], the large (40 meV) magnon gap measured for out-of-plane fluctuations was explained in terms of the above $J_{\rm H}$ induced anisotropy term $\Gamma_1$, although $J_{\rm H}$ was not explicitly mentioned in the very brief discussion, as the main focus was on resolving the small ($\sim$ 3 meV) magnon gap corresponding to the easy-axis anisotropy and basal-plane fluctuations via high-resolution RIXS and inelastic neutron scattering (INS). The proposed mechanism for the easy-axis anisotropy involve interlayer coupling and orthorhombic distortion which are beyond the scope of this work. 

%The explanation presented here is misleading as it does not identify $J_{\rm H}$ as the source of easy-plane anisotropy, while in the reference cited here (Khaliullin 2009) it is clearly mentioned that there is no true magnetic anisotropy in the absence of $J_{\rm H}$, and the given expression for anisotropic spin interactions is obtained when $J_{\rm H}$ is turned on.

For the easy-plane anisotropy, our conclusions are in agreement with the above analysis, including no true anisotropy in the absence of $J_{\rm H}$, and the tetragonal distortion term only affecting the magnitude and not being the source of anisotropy. In terms of symmetry breaking, the above $J_{\rm H}$ induced easy-plane anisotropy term $\Gamma_1$ obtained via strong coupling expansion is equivalent to our local anisotropic interaction term (dominantly $-2J_{\rm H} S_{1z} S_{3z}$) derived using the transformation. The major difference is that while in Ref. [17] the $\Gamma_1$ value was treated as a fitting parameter in the 40 meV magnon gap analysis, in our work the explicit form of the anisotropic interaction term [$2J_{\rm H}(S_{1z}S_{2z}-S_{1z}S_{3z})$] has been derived, and the magnon gap is determined using the same microscopic Hamiltonian parameters which account for the high energy magnon feature (strong zone boundary dispersion) within a unified scheme. Furthermore, all physical terms are treated on the same footing, which is especially important for the weakly correlated $5d$ systems.\cite{iridate_one} In general, the nature of magnetic anisotropy depends, through the sign and magnitude of the $J$=3/2 sector moments $S_{2z}$ and $S_{3z}$, on the lattice, band structure, and magnetic ordering, which is of particular interest for the honeycomb lattice compounds $\rm Na_2IrO_3$ and $\alpha$-$\rm RuCl_3$.

\section{Conclusion}
While all Coulomb interaction terms are invariant under same pure-spin rotation for all three pure ($t_{\rm 2g}$) orbitals, the Hund's coupling and pair hopping interaction terms were shown to explicitly break pseudo-spin rotation symmetry systematically due to the spin-orbital entanglement. Transformation of the various Coulomb interaction terms to the pseudo-spin-orbital basis formed by the $J$=1/2 and 3/2 states therefore provides a physically transparent approach for investigating magnetic ordering and anisotropy effects in the perovskite ($\rm Sr_2 Ir O_4$) and other $d^5$ pseudo-spin compounds. Explicitly pseudo-spin symmetry-breaking terms were obtained (dominantly $-2J_{\rm H} S_{1}^z S_{3}^z$), resulting in easy $x$-$y$ plane anisotropy and magnon gap for the out-of-plane mode, highlighting the importance of mixing with the nominally non-magnetic $J$=3/2 sector in determining the magnetic properties of $\rm Sr_2 Ir O_4$.

% \cite{iridate_two,hc_two}

\end{document}